**Double-percolation and magnetoresistance effects in ferromagnet-superconductor nanoparticle composites**

Xiangdong Liu[1], Raghava P. Panguluri[1], Daniel P. Shoemaker[2*], Zhi-Feng Huang[1], and Boris Nadgorny[1].

[1]Department of Physics and Astronomy, Wayne State University, Detroit, Michigan 48201, USA

[2]Materials Department and Materials Research Laboratory, University of California, Santa Barbara, CA 93106-5121, USA

## Abstract

We investigate transport and magnetotransport properties of binary networks composed of superconducting ($MgB_2$) and ferromagnetic ($CrO_2$ or $La_{1/3}Sr_{2/3}MnO_3$ (LSMO)) nanoparticles. While for the LSMO/$MgB_2$ system a single percolation threshold is observed, for $CrO_2$/$MgB_2$ binary composites an anomalously high resistance state with two distinct percolation thresholds corresponding to conductor–insulator and superconductor-insulator transitions is found. The existence of this double percolation effect becomes possible when the interface conductance between the two different constituents is suppressed and the condition for the two thresholds $p_c^{MgB_2} + p_c^{CrO_2} > 1$ is satisfied. For $MgB_2$ an unusually large value of the threshold $p_c^{MgB_2} = 0.78$ is observed, which can be explained by the significant geometric disparity between nanoparticles of the two constituents, resulting in a large excluded volume for $MgB_2$ nanoparticles. The scaling behavior near both thresholds is determined, with the two critical exponents identified: $\mu \approx 2.16 \pm 0.15$ for the insulating-conducting transition on the $CrO_2$ side and $s = 1.37 \pm 0.05$ for the insulating – superconducting transition on the $MgB_2$ side. We also measure the magnetoresistance for the entire series of $CrO_2$/$MgB_2$ samples, with a maximum of approximately 45% observed near the percolation threshold at liquid He temperatures.

## I. INTRODUCTION

Composite materials, which can enable new, intelligently tailored functionalities, are becoming increasingly important in numerous applications. These new functional materials might require various combinations of different constituents.[1] Many properties of these complex materials can be described by the classical percolation models, which serve as a powerful tool for understanding composite materials where geometric phase transitions appear.[2] Percolation effects governing various properties of these materials have been extensively investigated in various complex systems, such as metal-insulator[3] and superconductor/insulator[4,5] composites as well as polymer blends[6], polycrystalline materials[7], and ferromagnetic materials.[8]

However, there are many important questions beyond the conventional percolation theory that need to be addressed, such as the extent to which the shape, size and the nature of conductance between constituent particles in a composite network influence its percolation threshold and scaling exponents. Moreover, for nanoparticle tunneling the geometric properties and electrical connectivity of individual particles are often not the same. While in the case of single-valued two-particle resistance that results from tunneling conductance in a composite system, the universality is believed to be preserved, the most general case of percolation tunneling with variable coupling strength may be non-universal.[9] When interaction between constituents is present in a composite system, as in this paper, additional degree of freedom for the study of percolation becomes available, leading to a new, non-universal scaling behavior. Such is the case with magnetic nanoparticles, where spin dependent transport plays an important role, and the system behavior can be characterized by magnetoresistance as well.[8,10] Percolation in ferromagnetic/superconducting composites can be even more complicated as interaction can be both direct (via dipole field from magnetic nanoparticles) and indirect (via proximity effect).

To study some of these questions in detail we designed and fabricated a number of hybrid nanocrystalline composites by combining half-metallic ferromagnets and superconductor nanoparticles at different volume ratios. This allowed us to investigate both the percolation effects and the electrical and magnetotransport properties of the composites. In our previous work we described the observation of a new double percolation effect[11], manifesting itself as two independent percolative transitions in the composite system, one for superconducting $MgB_2$[12] and the other for half-metallic[13] $CrO_2$. Correspondently, we identified two percolation thresholds for the conductivity of each component; one of them (for $MgB_2$) was anomalously high. We argued that the constituent particle geometric shape and size played a crucial role in achieving the necessary conditions enabling the observation of the double percolation effect. This is confirmed by replacing the ferromagnetic $CrO_2$ particles by roughly spherical $La_{1/3}Sr_{2/3}MnO_3$ (LSMO) particles[14]; a much lower threshold then is found. In this paper, we provide a compendium of data on these composite systems, which include not only the electrical transport but also magnetotransport measurements and discuss the observed effects in detail.

## II. EXPERIMENTAL DETAILS

A series of composite ferromagnetic/superconducting samples were fabricated by using the traditional cold pressed technique. Chromium dioxide ($CrO_2$) and $La_{1/3}Sr_{2/3}MnO_3$ (LSMO) powders were the two different ferromagnetic components, while magnesium diboride ($MgB_2$) was the common superconducting component. In our experiments we used commercial $CrO_2$ and $Mg_2$ powders, whereas LSMO powders were obtained from sintered crystallized pellets. For the latter, hydrated acetates of manganese, lanthanum, and strontium were mixed in de-ionized water at 80°C and stirred until the liquid had evaporated. The resulting powder was calcined in air at 1000°C for 1 hour, then allowed to cool to room temperature and pressed into pellets at 17

MPa. The calcined pellets were sintered at 1500°C for 16 hours with heating and cooling rates of 10 and 20C/min, respectively. High firing temperatures (> 1200°C) were required to achieve proper oxygenation and a metal-insulator transition near room temperature. Powder X-ray diffraction (Philips X'Pert with Cu-Kα radiation) was performed and Rietveld analysis using the XND software package[15] confirmed the sample to be phase-pure $La_{0.3}Sr_{0.7}MnO_3$. The two ferromagnet/superconductor components were mixed according to their weight ratios; the mixed powder was then ground in a glass mortar for one hour. As the surface of all of the constituents can be easily oxidized, in order to obtain reproducible results it was necessary to perform this process in a controlled environment, with the humidity kept below 30%. The ground powder was then compressed at 10 GPa in a hydraulic press into 5mm diameter and approximately 1mm thick pellets. Immediately after fabrication four gold wires were attached to the pellet with silver paste. The sample was then mounted on a Physical Property Measurement System (PPMS) puck for transport and magnetotransport measurements. The PPMS excitation current of 2μA (with the voltage limit of 5mV) was used for most resistance measurements. The pellet resistance was measured in the temperature range from 300K to 2K; the magnetoresistance (MR) was measured in the magnetic fields of up to 5T. A total of 23 samples were prepared according to the weight fraction $x$ of $CrO_2$: $x$ = 0, 0.1, 0.25, 0.29, 0.31, 0.32, 0.33, 0.34, 0.35, 0.36, 0.365, 0.37, 0.375, 0.38, 0.39, 0.40, 0.41, 0.42, 0.45, 0.5, 0.6, 0.8 and 1. Using the bulk densities of $CrO_2$, $\rho_1$=4.89 g/cm$^3$, and of $MgB_2$, $\rho_2$ = 2.57 g/cm$^3$, weight fractions $x$ can be converted to volume fractions $p$:

$$p = \frac{x/\rho_1}{x/\rho_1 + (1-x)/\rho_2} = \frac{1}{1+(\rho_1/\rho_2)(1/x-1)} = \frac{1}{1+1.903(1/x-1)}. \qquad (1)$$

In order to measure the packing fractions of the composites, pure $CrO_2$ and $MgB_2$ pellets as well as a composite pellet (near the percolation threshold of $MgB_2$) were prepared. The packing fraction was determined from the known masses, volumes and bulk densities of $CrO_2$, LSMO, and $MgB_2$. A different geometry was used to determine the interface properties and the contact resistance of $MgB_2/CrO_2$ particles. In the latter experiment, we first prepared a sample consisting of only one type of powder, cold pressing it using the same conditions and in the same geometry as used for other composite samples. Upon placing the other type of powder on top of the fabricated sample, we cold pressed them together one more time to form a pellet with a well defined interface between the two different components. Composite samples of different compositions were characterized by scanning electron microscopy (SEM). The average size and shape of the particles, as well as their size distribution, were determined from the analysis of the respective SEM images (see Fig.1). The spatial distribution of the two types of particles was monitored by Energy Dispersive X-ray Spectroscopy (EDS), as shown in Fig.2.

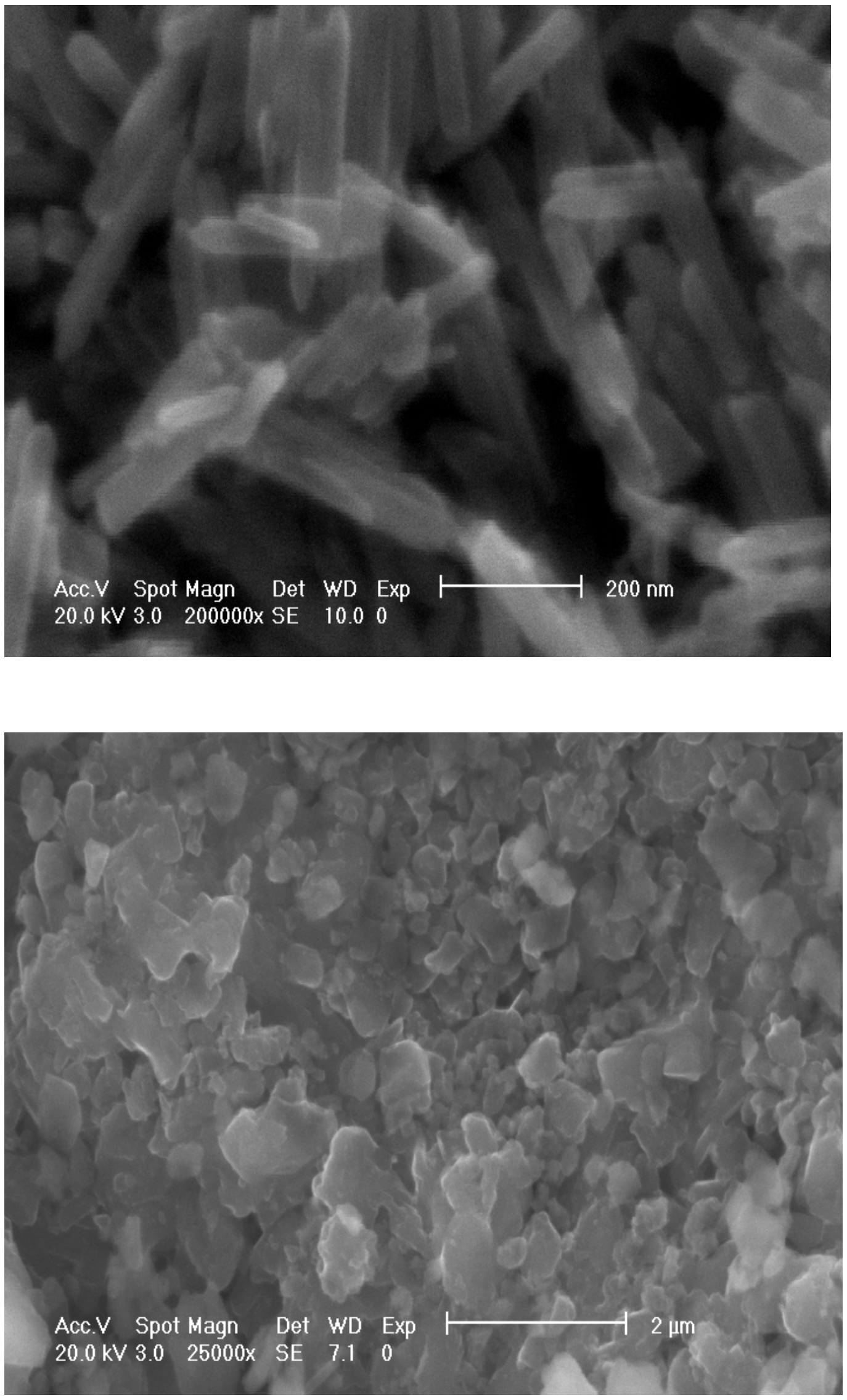

Fig. 1. Top: SEM image of $CrO_2$ particles; bottom: SEM image of $MgB_2$ particles

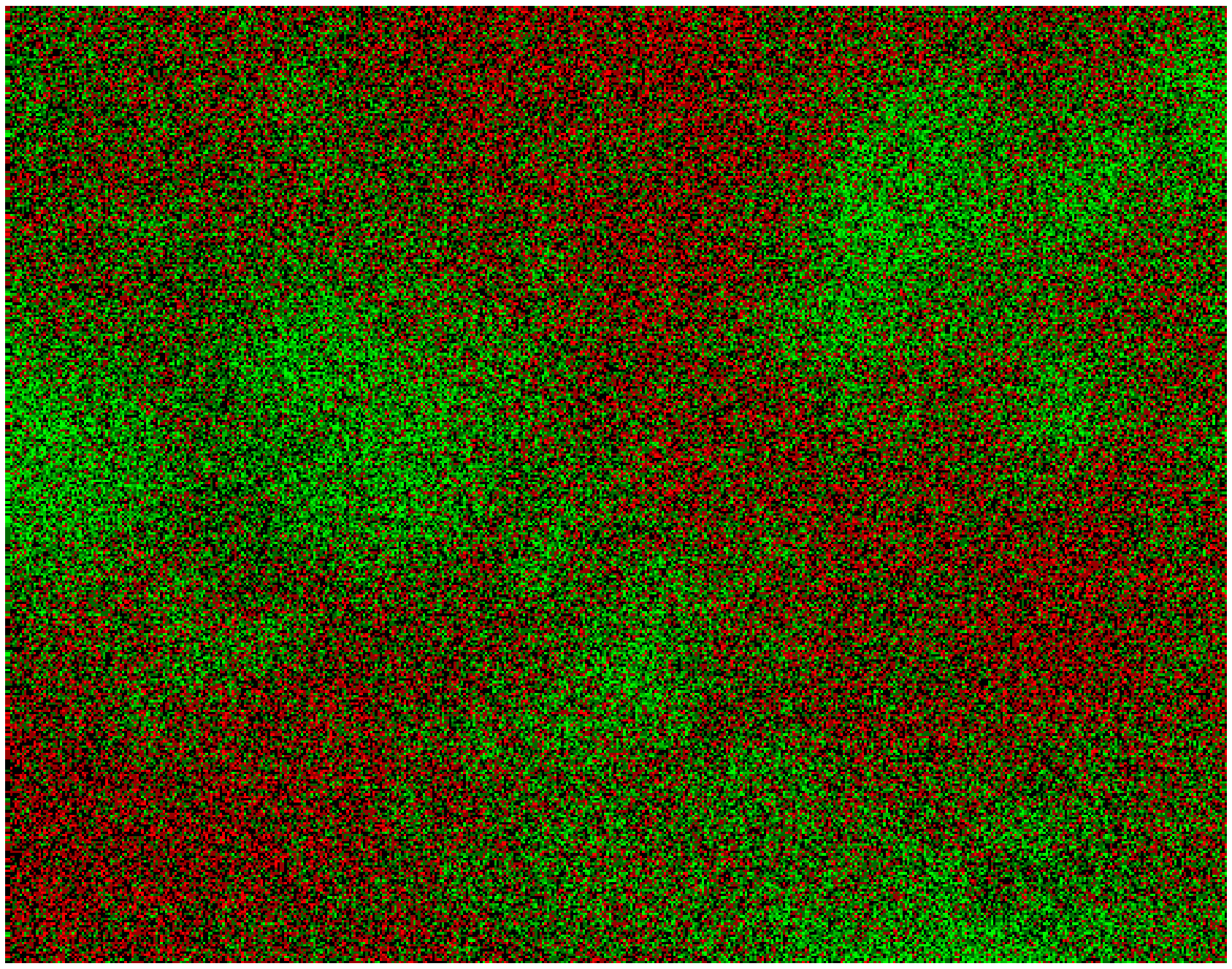

Fig. 2. (Color Online). EDS mapping of $(CrO_2)_{0.4}(MgB_2)_{0.6}$ sample (green: Mg; red: O).

The $MgB_2$ particles are approximately of spherical shape, with the average diameter $D$ of about 500 nm. In contrast, the $CrO_2$ particles are rod-like, with the typical length $L$ of about 300 nm and the width $d$ of about 40 nm. To obtain the particle distribution of binary composites, we use the SEM energy dispersive spectrometry for the mapping of elemental Mg and O. An example is shown in Fig. 2, for a composite sample of $x = 0.4$ (i.e., volume fraction $p = 0.26$), with the total particle packing density being about 81% (Note that the typical packing densities for pure particle samples are much less: ~ 46% for cold-pressed pure $CrO_2$ and ~62% for pure $MgB_2$ as we measured). The LSMO particles characterized by SEM were also approximately spherical with the polydispersed size distribution ranging from 1 to 5 μm. $(LSMO)_x(MgB_2)_{1-x}$ composite samples were fabricated by the same technique as $(CrO_2)_x(MgB_2)_{1-x}$ samples.

## III. RESULTS AND DISCUSSIONS

### A. Double percolation effect

In order to study the percolative transition, electrical characterization was conducted for the samples with variable compositions. First, several samples of basic configurations have been tested, including a pure $CrO_2$ sample, a pure $MgB_2$ sample, and a sandwich-like sample composed of one $CrO_2$ and one $MgB_2$ layer separated by an interface. The corresponding $R$ vs. $T$ curves are shown in Figs. 3(a), 3(b), and 3(c) respectively. The maximum resistance of the pure $CrO_2$ sample is ~ 55Ω at 2K; the maximum resistance of pure $MgB_2$ is 0.04 Ω at room temperature with the superconducting transition observed at $T_C$ ~ 37K. On the other hand, the sample with the interface junction of $MgB_2/CrO_2$ has the maximum resistance approaching 1000 Ω at 2K, approximately 20 times higher compared to the pure $CrO_2$ sample over the whole temperature range (the resistance of the pure $MgB_2$ sample above $T_C$ is negligible). This result indicates that the interface resistance between *two different types* of particles ($CrO_2/MgB_2$) is orders of magnitude higher than that between *the same types* of particles ($CrO_2/CrO_2$ or $MgB_2/MgB_2$), which has important implications for further analysis and discussion.

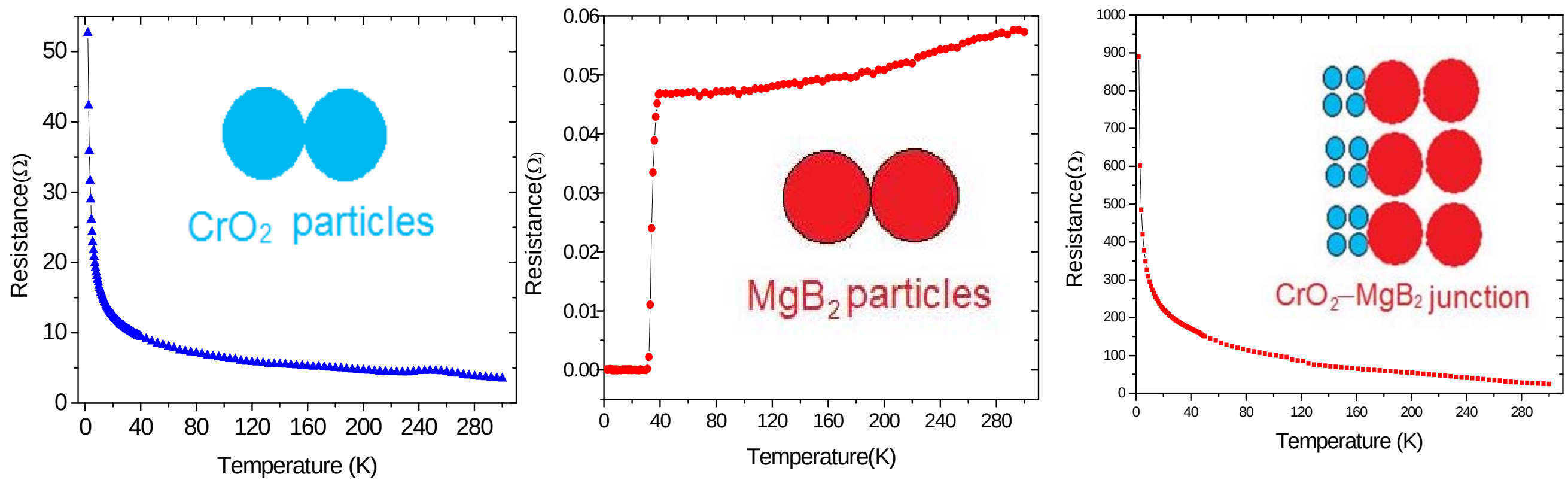

Fig.3. Temperature dependent resistance: Left: cold-pressed $CrO_2$ sample; Center: cold-pressed $MgB_2$ sample; Right: cold-pressed sandwich-like sample with a $CrO_2$/ $MgB_2$ junction.

Having obtained these results, we focus on samples of cold-pressed random composites $(CrO_2)_x(MgB_2)_{1-x}$. Results of the measurements for the temperature dependent resistance of various composition are shown in Fig. 4. The plots of sample resistance can be divided into two types: for the samples with high $MgB_2$ concentration, their *R* vs. *T* relation is similar to that of $MgB_2$, i.e. they are either superconducting or have very low resistances below a certain temperature, with the resistance generally increasing as the $MgB_2$ volume fraction decreases (see the first seven panels in Fig. 4).

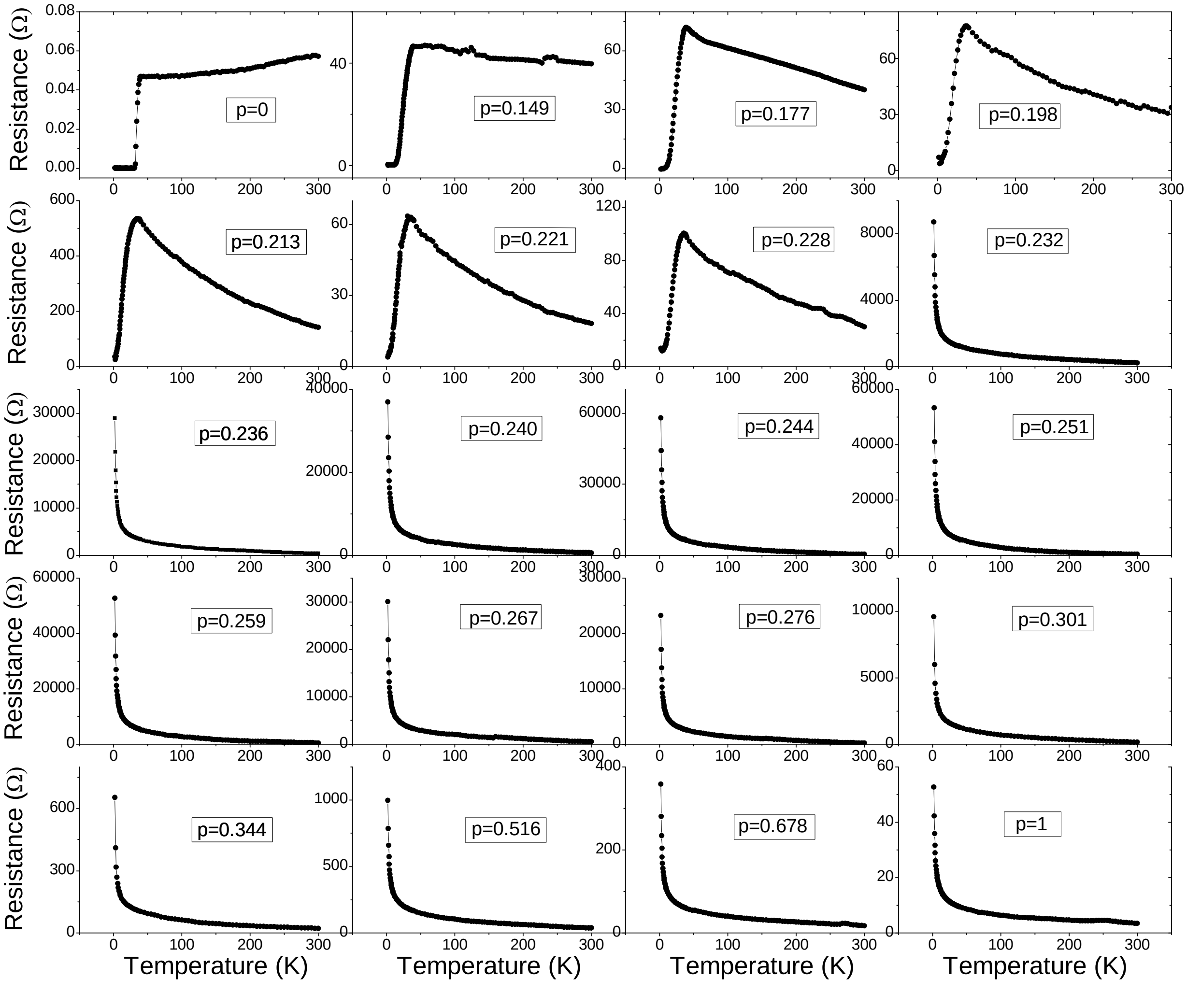

Fig. 4. Resistance as a function of temperature for 20 different composites with volume fractions *p* ranging from 0 to 1.

Samples with higher $CrO_2$ concentration have $R$ vs. $T$ dependence similar to that of $CrO_2$ (see the last 13 panels in Fig. 4). Interestingly, the resistance does not increase monotonically as the $CrO_2$ volume fraction increases, as shown in detail in Fig. 5 where the measured electrical resistance for the samples of various compositions (volume fractions $p$) is plotted at different temperatures. While the resistance on the $MgB_2$-rich side ($0 < p < 0.22$) is very low and the resistance on $CrO_2$-rich side ($0.34 < p < 1$) is also fairly low, the resistance of the composite samples attains peak values within a narrow intermediate range of $0.22 < p < 0.34$. At low temperatures the maximum value of the resistance is about *three orders of magnitude higher* than that of the pure $CrO_2$ sample; e.g., at 2K the maximum resistances are 450 Ω for $p < 0.22$, 900 Ω for $p > 0.34$, but close to 55 kΩ around $p = 0.25$.

We can attribute this unusual behavior to a new double percolation effect appearing in this type of nanocomposite system. As we have shown in our most recent study[11], this involves the development of two separate uncorrelated percolative networks with conductive or superconductive paths, with different percolation thresholds for the two constituents: $p_c^{CrO2} \approx 0.34$ and $1 - p_c^{MgB2} \approx 1 - 0.78 = 0.22$. The difference of these two thresholds indicates that the corresponding percolation transitions are not symmetric, a direct consequence of significant asymmetry in the size and shape of the constituent particles that will be discussed below. Note that there is a small but reproducible resistance peak in a very narrow volume fraction range close to the transition regime of $MgB_2$ (see the inset of Fig. 5), which is absent on the $CrO_2$ side of the transition. At the same time, the $MgB_2$ side exhibits a much sharper resistance transition compared to that of $CrO_2$.

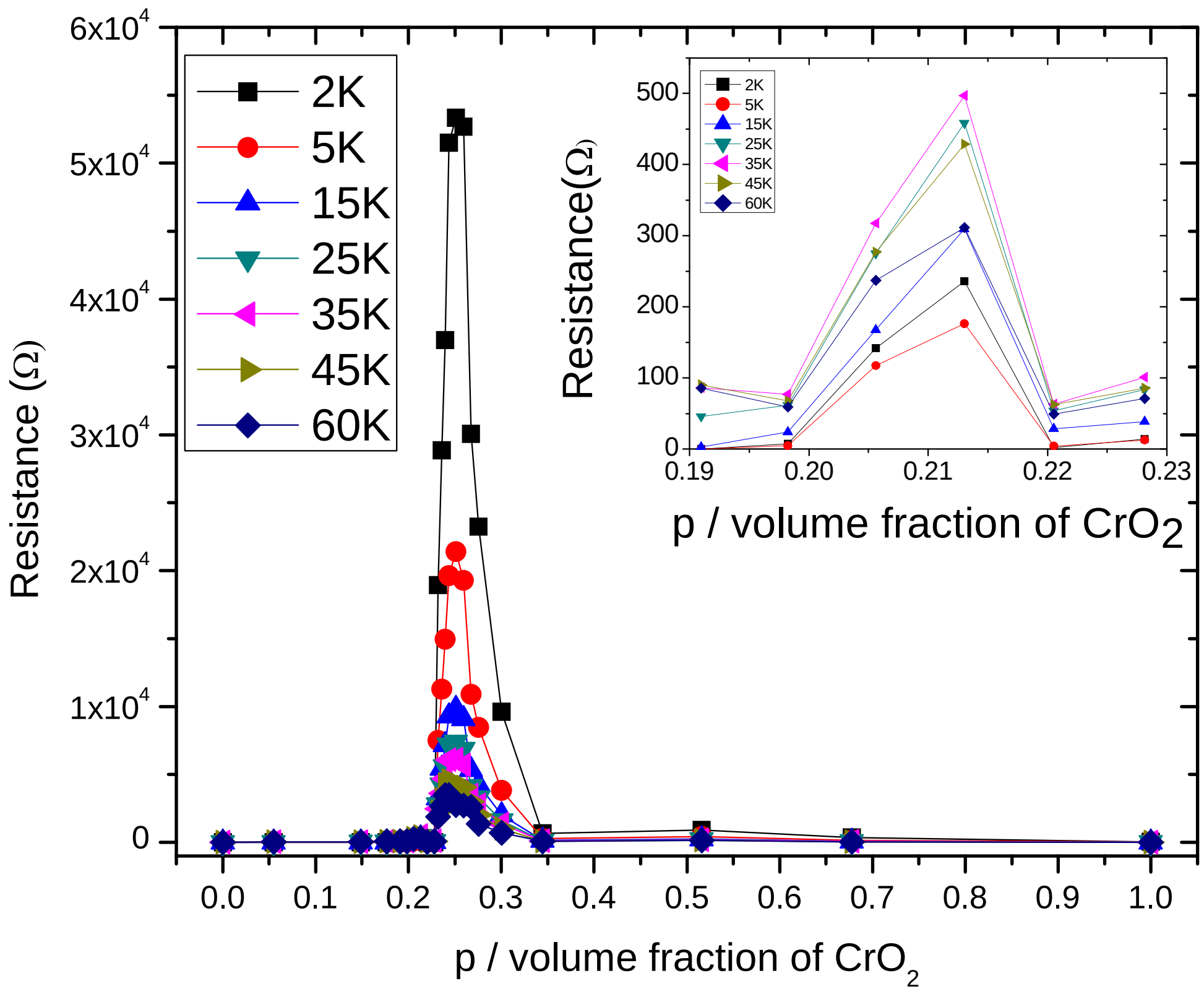

Fig. 5. Resistance of $(CrO_2)_p(MgB_2)_{1-p}$ cold pressed composites as a function of the volume fraction $p$ of $CrO_2$ at temperatures ranging from 2K to 60K. Inset: The same $R$ vs. $p$ curves in the enlarged region of $0.19 < p < 0.23$ which is close to the percolation threshold of $MgB_2$.

In order to explain these surprising results, we recall that the resistance of a heterogeneous interface (i.e., $MgB_2/CrO_2$) is orders of magnitude larger than that of homogeneous ones (i.e., $CrO_2/CrO_2$ or $MgB_2/MgB_2$), particularly at low temperatures (see Fig. 3). Therefore, it is reasonable to assume that in the samples of random compositions the path maximizing the overall interface region between $MgB_2$ and $CrO_2$ clusters would maximize the sample resistance. Note the two independent percolative transitions observed, one from an insulating state to a conductive state (on the $CrO_2$–rich side) and the other from an insulating state to a superconductive state (on the $MgB_2$-rich side), which are realized at certain compositions of $CrO_2$ (or $MgB_2$). If, in the range of intermediate compositions, none of the two constituent clusters percolate (which, as we will discuss below, requires certain conditions to be fulfilled), one can expect the maximum resistance somewhere in this range (as seen in Fig. 5), and thus the double percolation effect.[11]

In a system exhibiting a percolative transition, the resistance near the percolation threshold can be expressed, in accordance with the percolation theory, by a power law:

$$R \sim \left| p - p_c \right|^{-\mu}, \qquad (2)$$

where the critical exponent $\mu$ depends on the dimensionality and the intrinsic conducting property (i.e., conducting or superconducting) of the system. For three-dimensional (3D) networks, as in our case, the critical exponents of conductor-insulator transitions are given by $\mu \approx 2$ for lattice percolation [16] and $\mu \approx 2.38$ for continuum percolation (the Swiss-cheese model),[17] while for a conductor-superconductor percolative transition the expected value of the exponent (typically denoted as $s$ instead of $\mu$) is 0.75.[18] To demonstrate that the two transitions in our composite systems are governed by the above scaling law, we present the log-log plots of resistance vs. volume fraction for both transitions, at $p_c^{MgB_2} = 0.78$ for the $MgB_2$ side and at $p_c^{CrO2} = 0.34$ for the $CrO_2$ side, in Figs. 6(a) and 6(b) respectively. A close examination of the data confirms that for both $MgB_2$- and $CrO_2$- dominated networks the transitions are percolative, with the average exponent $\mu \approx 2.16 \pm 0.15$ (for the insulating-conducting transition on the $CrO_2$ side) and $s = 1.37 \pm 0.05$ (for the insulating – superconducting transition on the $MgB_2$ side) as identified from Figs. 6(a) and 6(b) respectively. Note that the standard deviation for the critical exponent is appreciably higher for the $CrO_2$ side and there is some weak temperature dependence of $\mu$, which varies from 1.97 at 60K to 2.36 at 2K. This may be related to the fact that conductance across the $CrO_2$ percolative network, especially at the lowest temperatures is dominated by tunneling. The experimentally obtained exponent on the $CrO_2$ side is very close to the theoretical values (which are between 2 and 2.38 for 3D lattice and continuum network respectively). On the other hand, the exponent on the $MgB_2$ side deviates significantly from the known theoretical result ($s \approx 0.75$) for the conductor-superconductor transition.[18]

The question of whether the critical exponent $s$ measured in our system should coincide to the one in a conductor - superconductor transition, is non-trivial. Indeed, the conductor - superconductor transition can be modeled by short circuiting an element in the resistor model,[19] while adding a link between the elements would correspond to the metal-insulator transition (on the metal side). However, in our system, it is impossible to separate these two operations. Additionally, the critical currents across the $MgB_2$/$MgB_2$ interface are likely to be nonuniform and determined not by the properties of the bulk $MgB_2$, but rather by the properties of these interfaces, i.e. the weak links between the $MgB_2$ particles (although these "weak" links are actually quite strong[20]). In other words, the superconducting cluster would propagate via the proximity/Josephson coupling between grains, which assumes that the metal-insulator percolative transition should occur first. This is consistent with our measured exponent ($s = 1.37$), which deviates from the conductor-superconductor transition value towards that of the conductor-insulator transition. Moreover, it is entirely possible that magnetic $CrO_2$ may affect the superconductivity of $MgB_2$ nanoparticles near the percolation threshold, particularly in view of "surface aging" reported for some of the $MgB_2$ nanoparticles.[21] In other words, there could be a significant difference between superconductors percolating in a non-interactive media, and in a media with highly localized magnetic moments, as has been shown for $CrO_2$ particles.[22] If this indeed is the case, the critical exponent is likely to scale with the particle size in some nontrivial way, particularly when the interaction length becomes comparable with the nanoparticle size. Another difference, compared to the previously reported results, is the location of percolation thresholds in the heterogeneous nanocomposites studied here. While the observed percolation threshold for $CrO_2$ ($p_c$ = 0.34) was close to the known theoretical limit, the threshold for $MgB_2$ ($p_c$ = 0.78) was significantly higher than expected. This puzzling result, which is strongly dependent on the geometric shape and size disparity of the constituent particles, is the key to understanding the double percolation effect, as we will describe below.

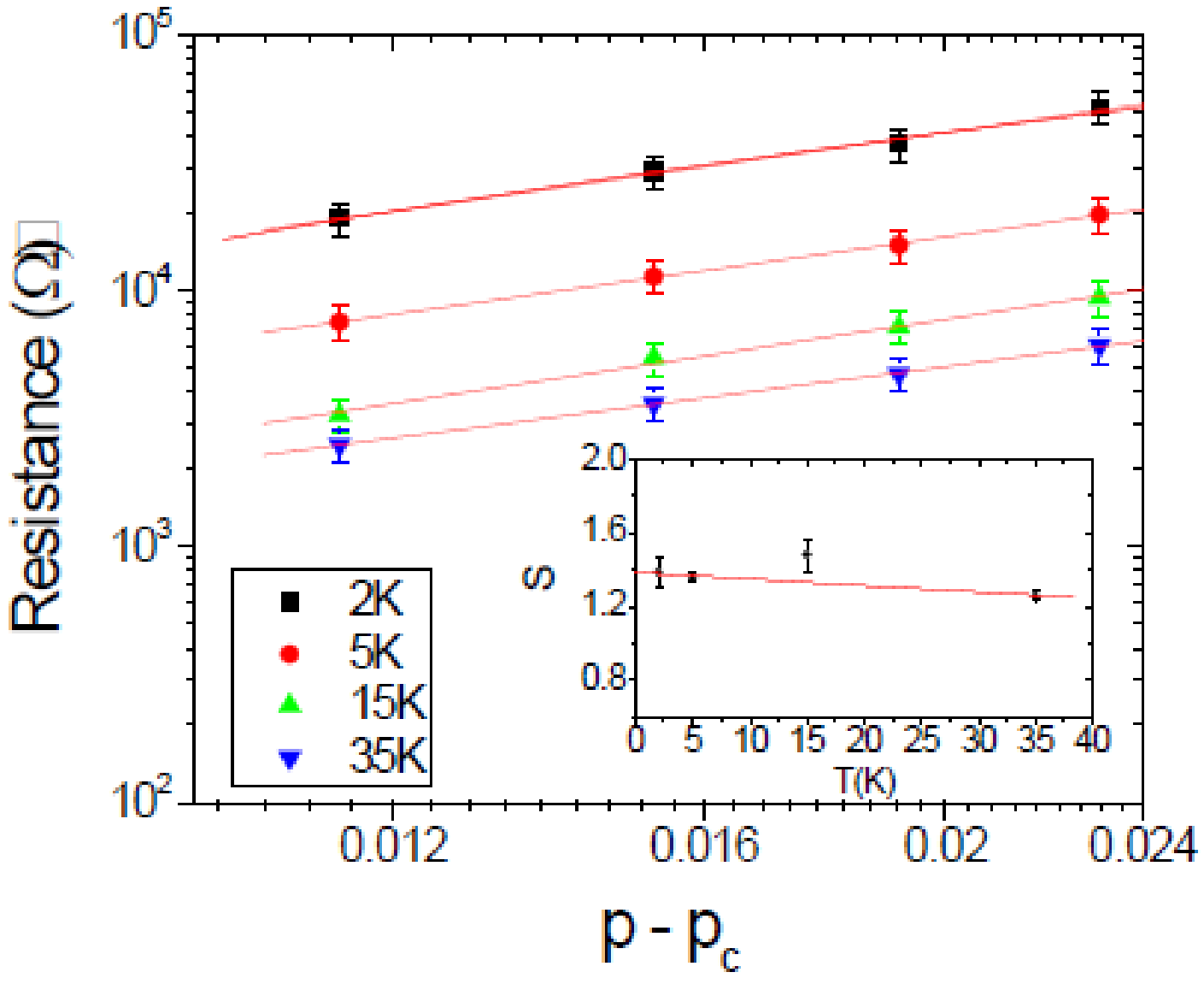

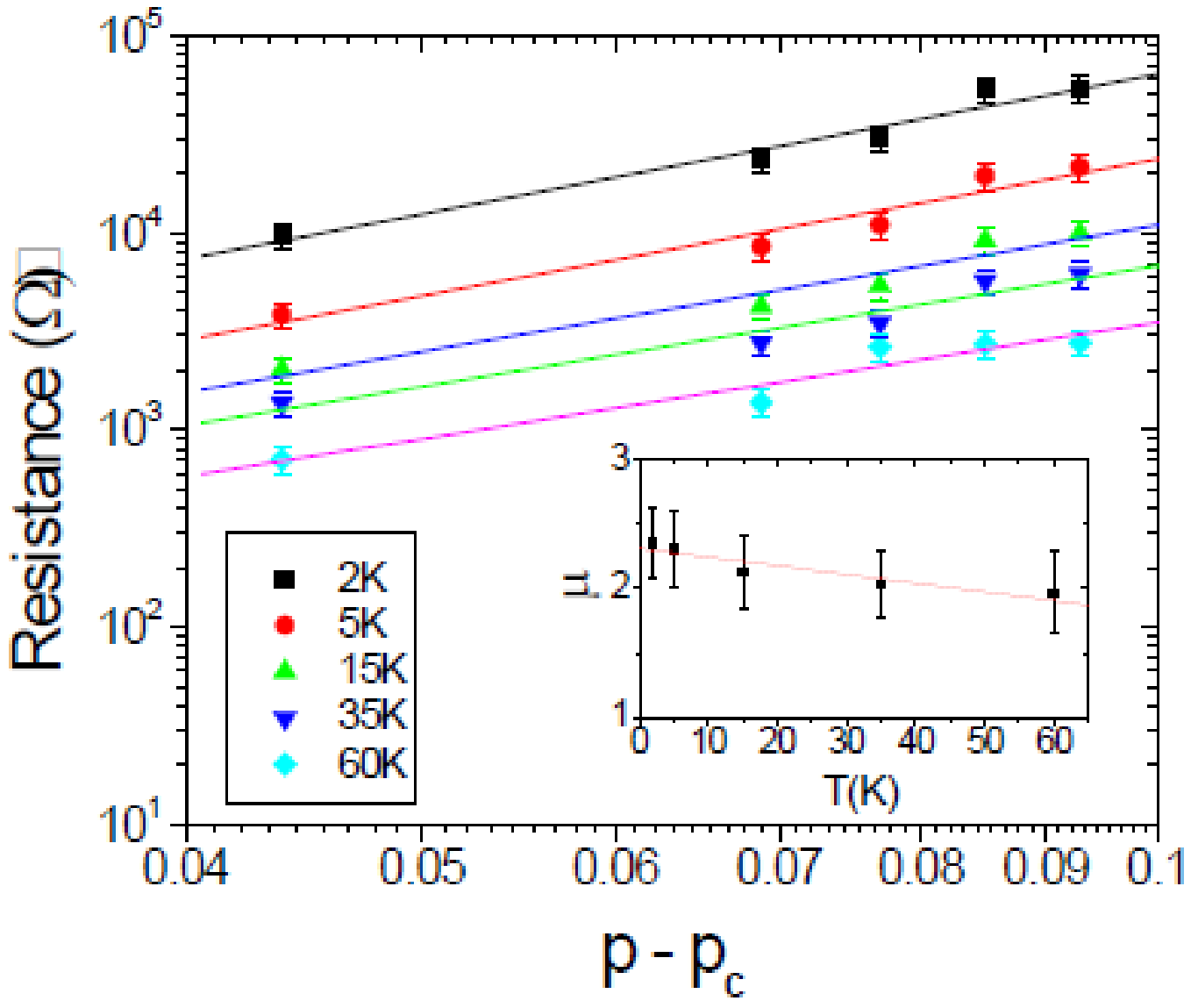

Fig. 6. Resistance scaling and the critical exponents, $s$ and $\mu$ (see the insets) near the percolation thresholds at different temperatures for the superconductor - insulator and conductor - insulator transitions. Top: $MgB_2$, $p_c = 0.78$; bottom: $CrO_2$, $p_c = 0.34$.

As determined from the SEM images, the length of the individual $CrO_2$ rod-like particles is only about one half the size of the $MgB_2$ particles; they are also very thin, with the length-to-width aspect ratio about 7.5:1, so that their volume is only about 0.5% of that of $MgB_2$ spheres. Thus the $CrO_2$ particles can be effectively treated as if they are placed in a continuum matrix, with the observed threshold $p_c^{CrO_2} \sim 0.34$ consistent with the known theoretical results[2] (e.g., $p_c \approx 0.31$ for site percolation in simple cubic lattice, and $p_c \approx 0.28$ for 3D continuum percolation model).

On the other hand, for $MgB_2$ particles the situation is more complicated. First, one has to take into account the packing density at the threshold, which as we measured is approximately 81.1% (for all the constituents, $MgB_2$ and $CrO_2$). Since $p_c^{MgB_2} = 0.78$ is the volume fraction, we need to convert it to the space occupation or probability fraction,[23] $P_c = f\,p_c$, where $f$ is the filling factor. Taking the experimental values, $f$ =0.811, $p_c$ = 0.78, we obtain $P_c = f\,p_c = 0.633$, which is within the limit of 0.64, the random close packing limit of hardcore spheres embedded in a continuum medium.[24] While this number is more reasonable than 0.78, which is obtained based purely on the volume fraction ratio, our samples are cold-pressed, further reducing the average distance between the adjacent particles. Hence the value of $P_c$ is expected to be much lower than the random packing limit and around the range of 0.2 – 0.5 as found in previous experiments of $MgB_2$ [25] and $YBa_2Cu_3O_7$ [26] superconductor systems or granular metal films.[27] To address this discrepancy, we need to examine the detailed configuration of packing between particles of large geometric contrast, as featured in our binary system. As illustrated in Fig. 7, voids between large $MgB_2$ spheres are very effectively filled by small $CrO_2$ sticks that are short and ultra-thin, particularly under the cold-pressing experimental condition, leading to the high packing density of $f$ = 0.811. Meanwhile, these filled $CrO_2$ sticks can very effectively screen, or cage larger $MgB_2$ spheres, as also seen from the calculation of excluded volume[26] in a sphere-stick system.[28] According to Onsager's result for two spherocylinders with diameters $D$ and $d$ and lengths $l$ and $L$, both randomly oriented[29], the excluded volume is given by

$$V_{ex} = \frac{\pi}{6}(d+D)^3 + \frac{\pi}{4}(d+D)^2(l+L) + \frac{\pi}{4}(d+D)lL \,. \qquad (3)$$

For the sphere-stick system considered here, one of the spherocylinders in Eq. (3) is a sphere ($MgB_2$) with $l = 0$ and diameter $D$, while the other is a thin stick ($CrO_2$) with length $L$ and diameter $d << D$, as shown in Fig. 7. Thus from Eq. (3) the excluded volume of this binary composite is simplified as

$$V_{ex} = \frac{\pi}{6}D^3 + \frac{\pi}{4}D^2 L \,. \qquad (4)$$

Our measurement of this $MgB_2/CrO_2$ system yields $L \approx D/2$ (see Sec. II), and thus $V_{ex}/V_{MgB2} \approx 7/4$, where $V_{MgB_2} = \pi D^3/6$ is the volume of $MgB_2$ spheres. The percolation threshold for $MgB_2$ can be approximated as:

$$P_c = \frac{N_c V_{MgB_2}}{V_{sys}} \qquad (5),$$

where $V_{sys}$ is the total volume of the system. From the simplest model,[30] we can estimate the number of spheres ($MgB_2$) at the percolation threshold as $N_c \approx V_{sys}/V_{ex}$. Substituting this estimate into Eq. (5), we obtain $P_c \cong \frac{V_{MgB_2}}{V_{ex}} \cong 4/7 = 0.571$, which is consistent with our experimental value of $P_c = 0.633$. Note that the percolation threshold is very sensitive to the particle shape and size. For example, paper-thin rectangular particles with a volume fraction close to zero would be extremely efficient in caging the spheres, with the threshold approaching 1.

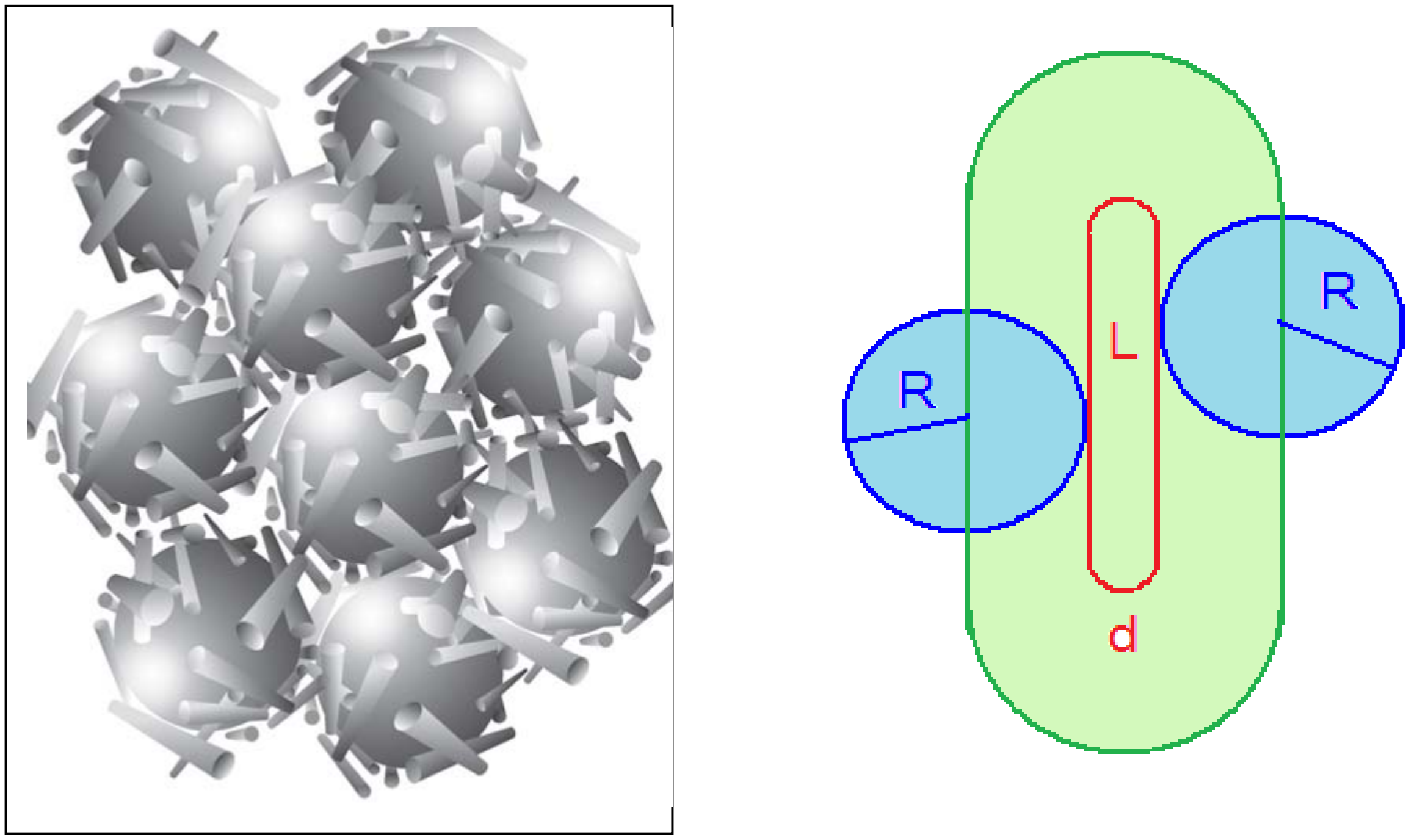

Fig. 7. Left: Schematics of $MgB_2$ spherical particles caged by $CrO_2$ thin rods; Right: Excluded volume of a system composed of a sphere (e.g., $MgB_2$ particles) and a thin rod (e.g., $CrO_2$ particles).

Since our explanation of the anomalously high percolation threshold for $MgB_2$ based on the concept of exclusion volume is essential for understanding our results, we have performed several additional experiments to confirm that the value of the threshold for $MgB_2$ is, indeed, independent of the double percolation effect *per se.* Two additional systems have been examined: $Cr_2O_3/MgB_2$ and LSMO/$MgB_2$. First, we annealed $CrO_2$ particles under vacuum using the recipe of Ref. 9 to convert them to $Cr_2O_3$, and study a $Cr_2O_3/MgB_2$ system. While for this system we no longer observe double percolation (as $Cr_2O_3$ particles are insulating), the value of the threshold for $MgB_2$ is found to be practically unchanged (0.76 compared to 0.78), which confirms our assumption that it is, indeed, the specific large geometric contrast of our system that drives the threshold higher. In the second experiment we measured the LSMO/$MgB_2$ composite system, in which the shape and size of the LSMO particles were quite different from $CrO_2$, close to spherical but generally larger in diameter than $MgB_2$ particles. Again, for this system only one percolation threshold was observed (as the LSMO particles were insulating) with $p_c^{\mathrm{LSMO}} \approx 0.35$, which is quite different from the results obtained for the case of $MgB_2$ particles percolating through the $CrO_2$ matrix, but close to the $p_c^{\mathrm{CrO2}} \approx 0.34$ for the $CrO_2$ percolating network in the spherical $MgB_2$ matrix.

To summarize, in order to be able to observe the double percolation phenomenon, two conditions need to be fulfilled independently: 1) The resistance between different types of constituent particles needs to be orders of magnitude higher than that between the same types of particles. In this case the particles of one type can be treated as a percolating matrix for the particles of another type. In this work, this is achieved by combining superconducting and ferromagnetic particles. At the nanoscale, the effect of interface tunneling between individual particles of different types (i.e., $MgB_2/CrO_2$) would be dominant as now the thickness of the interface region is comparable to the individual particle size, leading to anomalously high resistance, particularly at low temperatures. 2) However, this is not enough - it is also necessary to have a concentration range in which none of the constituent networks can percolate. Therefore, it is expected that this effect will not be observed for spherical particles of approximately the same size, as for both types in our cold pressing conditions the percolations thresholds would be about 0.2-0.3 (as we have observed for spherical LSMO particles), and thus there will be no region where none of the constituents can percolate (i.e., the high resistance region where both networks are separated). However, in a more general case of a large shape and size contrast between constituent particles this phenomenon can be observed due to an anomalously high percolation threshold of one of the constituents. The condition for the double percolation effect to occur in a binary composite system of materials I and II can then be written in the form:

$$p_c^{\mathrm{I}} + p_c^{\mathrm{II}} > 1 \quad (5)$$

where $p_c^{\mathrm{I}}$ and $p_c^{\mathrm{II}}$ are the respective percolation thresholds. If the inequality holds, then there is a region in which neither network I nor II can percolate, resulting in the double percolation effect. We note that, while in 2D this condition is relatively easily satisfied (with $p_c > 0.5$ in many 2D systems of various lattice symmetry[2]), in 3D it is generally not expected and can only be satisfied if the two types of particles have dissimilar geometries as we have shown here in the system of $MgB_2/CrO_2$ nanocomposites.

**B. Magnetoresistance effects**

To further explore the structure-property relation in this composite system, we have measured the sample conductivity under various conditions of magnetic fields, and observed two types of magnetoresistance (MR) effects in the $MgB_2/CrO_2$ compounds: Samples with $MgB_2$ volume fraction below the percolation threshold show negative MR, while positive MR is obtained for samples with $MgB_2$ volume fraction above the threshold. For the first type, the resistance of $MgB_2/CrO_2$ samples below the $MgB_2$ percolation threshold is measured reversibly in magnetic fields between -5T and 5T, as shown in Fig. 8 for composite $(MgB_2)_{0.39}(CrO_2)_{0.61}$ (the left panel). At high enough temperature (>15K), the resistance curves have the characteristic butterfly shape, as shown on the right panel of Fig. 8.

Our results can be compared with those obtained for half-metallic $CrO_2$ powders that were cold pressed in a matrix with antiferromagnetic $Cr_2O_3$ particles[8], where the characteristic butterfly-shape curves with reversible high-field slope were observed, with the two peaks occurring at the temperature dependent coercive field. This indicates that the magnetoresistance might be associated with the alignment of the magnetization of $CrO_2$ particles and the spin dependent tunneling[8].

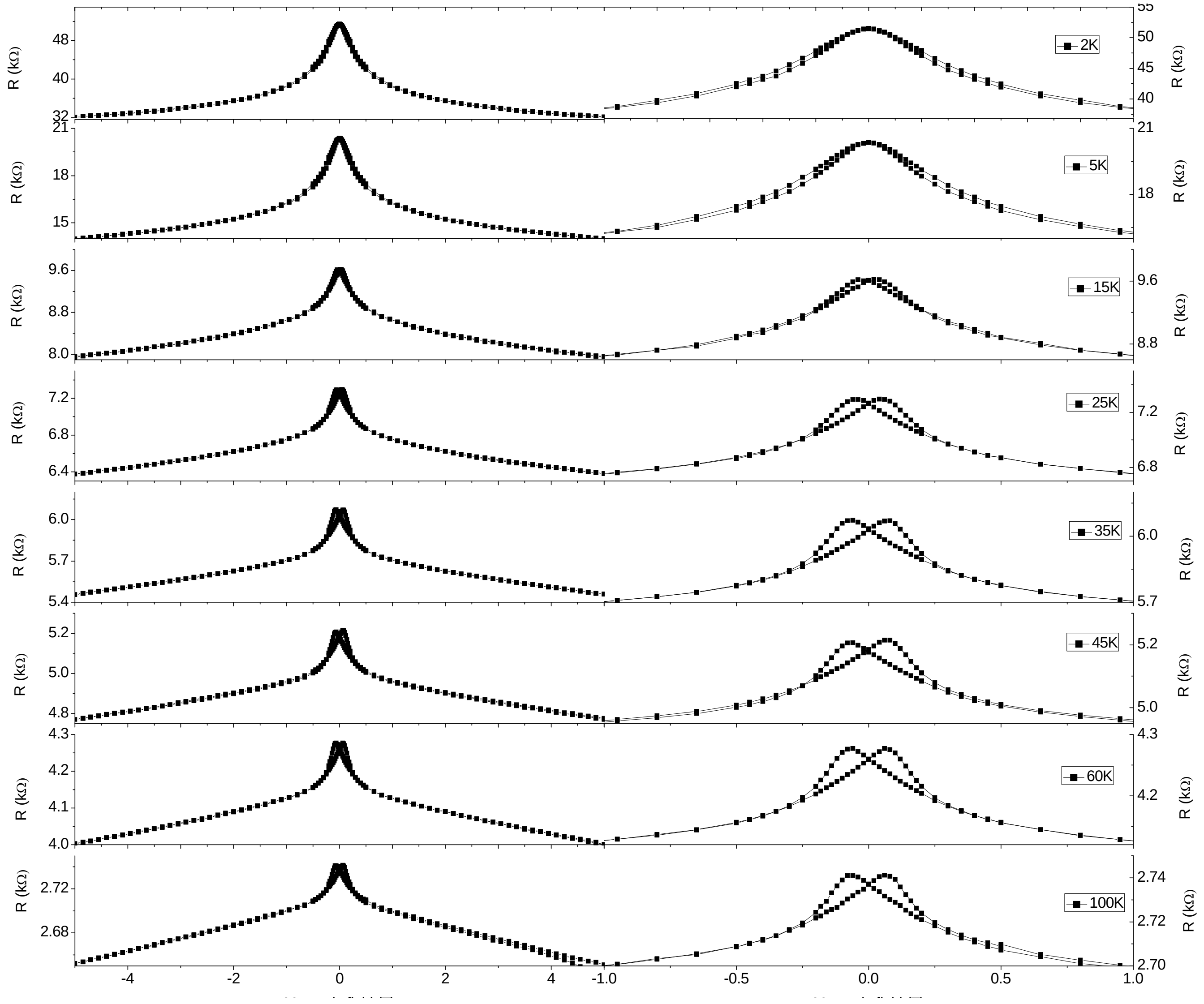

Fig. 8. Resistance of $(CrO_2)_{0.39}(MgB_2)_{0.61}$ cold-pressed sample (with the $MgB_2$ concentration below the percolation threshold) as a function of magnetic field at different temperatures. Left panel: full range of magnetic field is from -5T to 5T; right panel: blow up of the central region (from -1T to 1T), with the characteristic butterfly-shape magnetoresistance curves appearing above 15K.

For our $MgB_2/CrO_2$ system, the maximum magnetoresistance generally occurs at the lowest measurement temperature ($T$ = 2K), as seen in Figs. 9 and 10. However, in contrast to the results of Ref. 8 for $CrO_2/Cr_2O_3$ powders, the peaks of magnetoresistance in Fig. 8 do not seem to coincide with the coercivity (~0.1T) at temperatures below 35K (which is the critical temperature

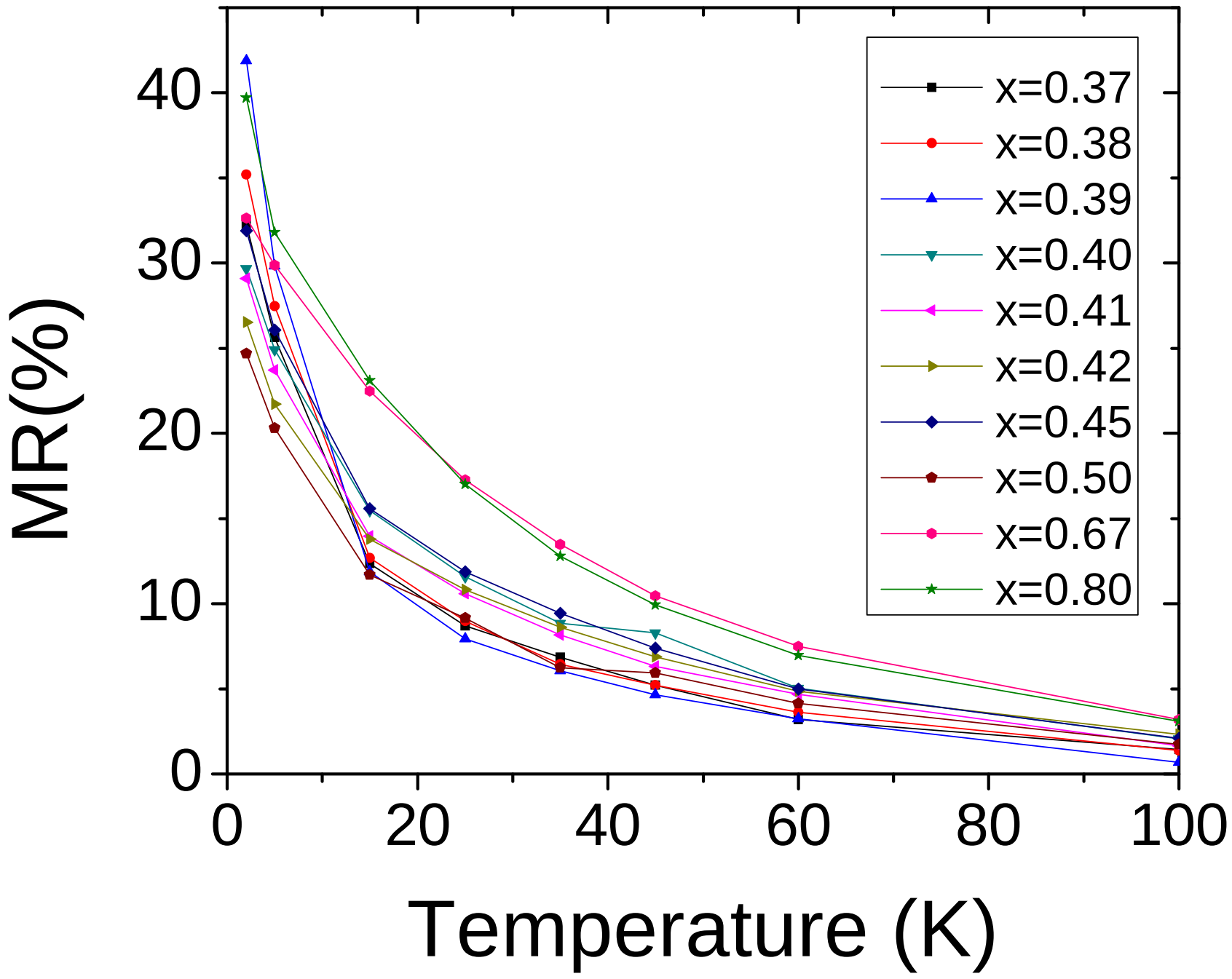

Fig.9. Temperature dependence of magnetoresistance for $(CrO_2)_x(MgB_2)_{1-x}$ with $x>0.36$.

$T_C$ of $MgB_2$). At temperatures below $T_C$ the position of the peaks shifts towards smaller fields (with the coercivity not changing or even slightly increasing in this temperature range), until they are hard to resolve at 2K and 5K. This suggests that the origin of magnetoresistance is likely to be different below and above the superconducting transition of $MgB_2$. The magnetoresistance as a function of temperature for different volume fractions of $CrO_2$ at different temperatures is plotted in Fig. 9, while the magnetoresistance as a function of volume fraction at different temperatures is plotted in Fig. 10. It can be seen from both plots that the magnetoresistance of this system has two maxima: one, MR = 42% at 2K, is obtained from the sample $(MgB_2)_{0.39}(CrO_2)_{0.61}$, which corresponds to the largest resistance of the system near the percolation threshold (see Fig. 5); the other of approximately the same value corresponds to a higher fraction of $CrO_2$, significantly above its percolation threshold, similar to the results of Ref. 8.

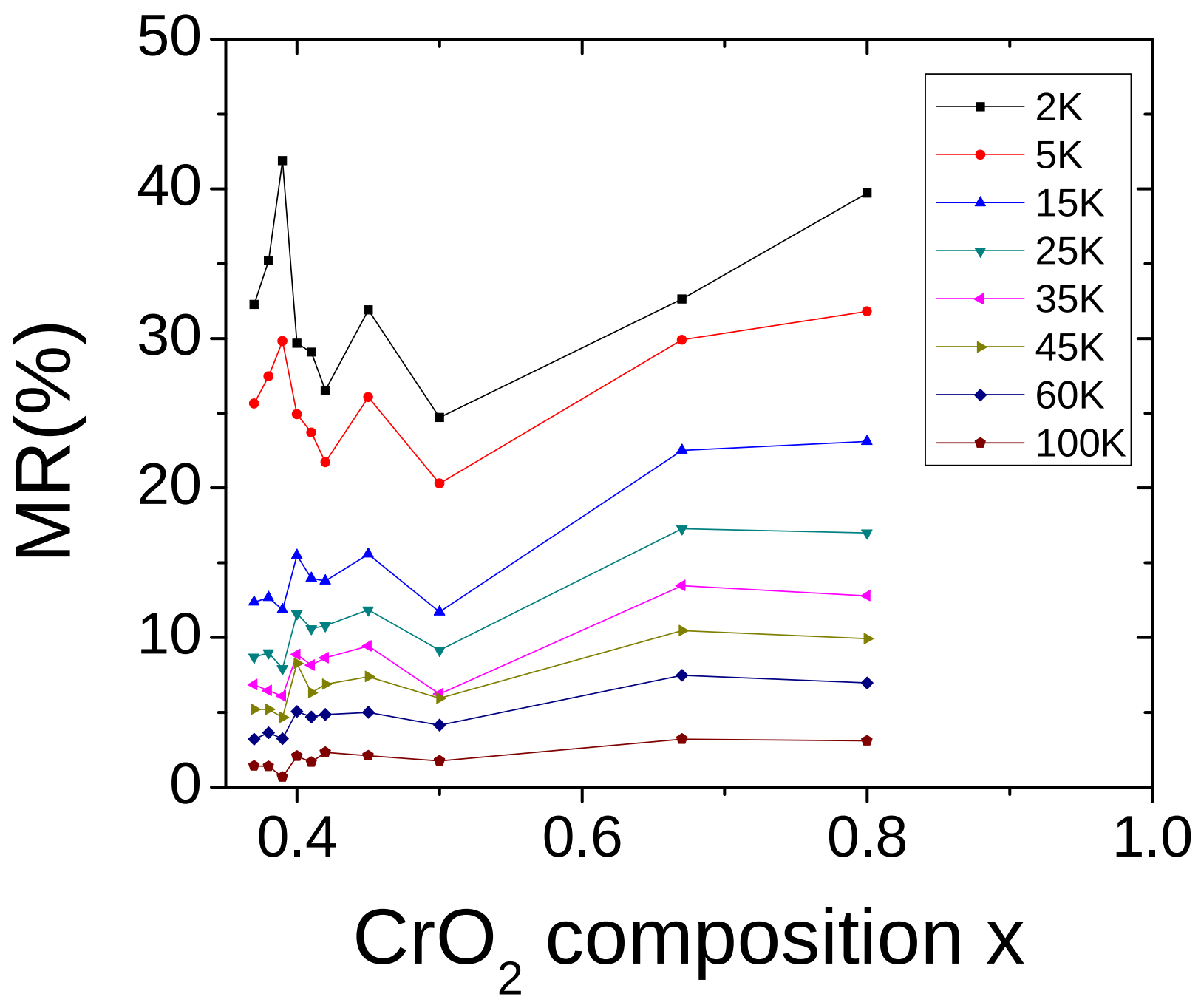

Fig. 10 Magnetoresistance vs. composition *x* for the samples with $x > 0.36$ at different temperatures.

The second type of magnetoresistance is observed in samples of which the $MgB_2$ volume fraction is above the $MgB_2$ percolation threshold. The typical magnetic field dependence of resistance is presented in Fig. 11. The resistance of the sample increases with increasing magnetic field at various temperatures, showing positive magnetoresistance. Interestingly, larger volume fraction of $MgB_2$ corresponds to higher maximum temperature at which the positive magnetoresistance is observed; e.g. for a pure $MgB_2$ sample, positive magnetoresistance is observed up to 37K, while for the low $MgB_2$ concentration samples positive MR is only observed below 25K, which may be related to the influence of the magnetic impurity phase ($CrO_2$).

A similar phenomenon of positive MR has been observed by Barber and Dynes,[31] who investigated granular Pb films near the insulator-superconductor transition in a magnetic field. They described three regimes corresponding to this behavior. At magnetic fields higher than some crossover point $H_x$, a power-law behavior $R \sim H^{\alpha}$ was found for intermediate fields, while at higher magnetic fields a logarithmic behavior $R \sim \log H$ was observed. The authors related the crossover field $H_x$ to the size of the regions of superconducting phase coherence, which increases with average thickness and the coupling between individual grains. In our case, a similar scenario might take place, that is, more $MgB_2$ particles are needed to realize global coherence.

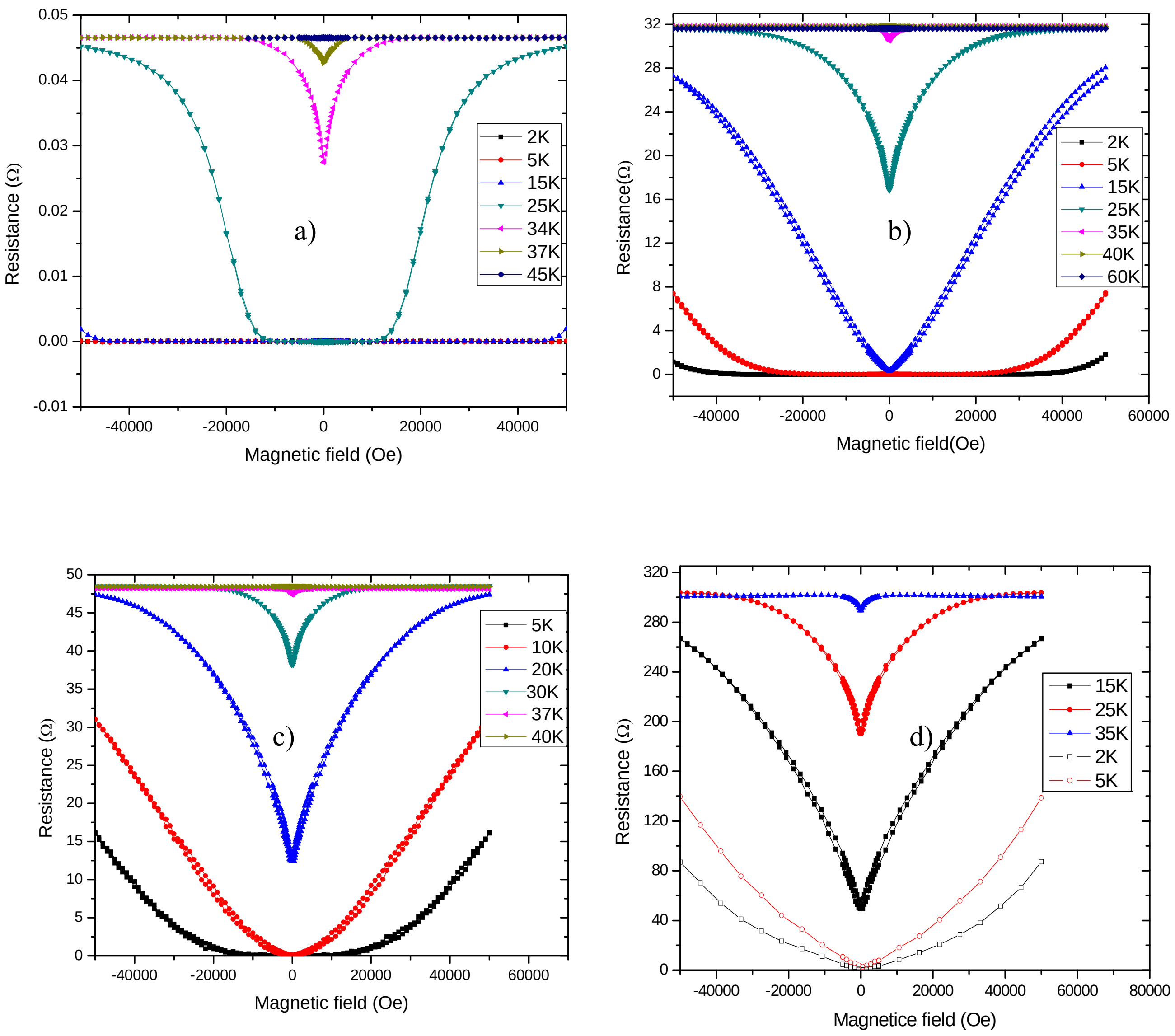

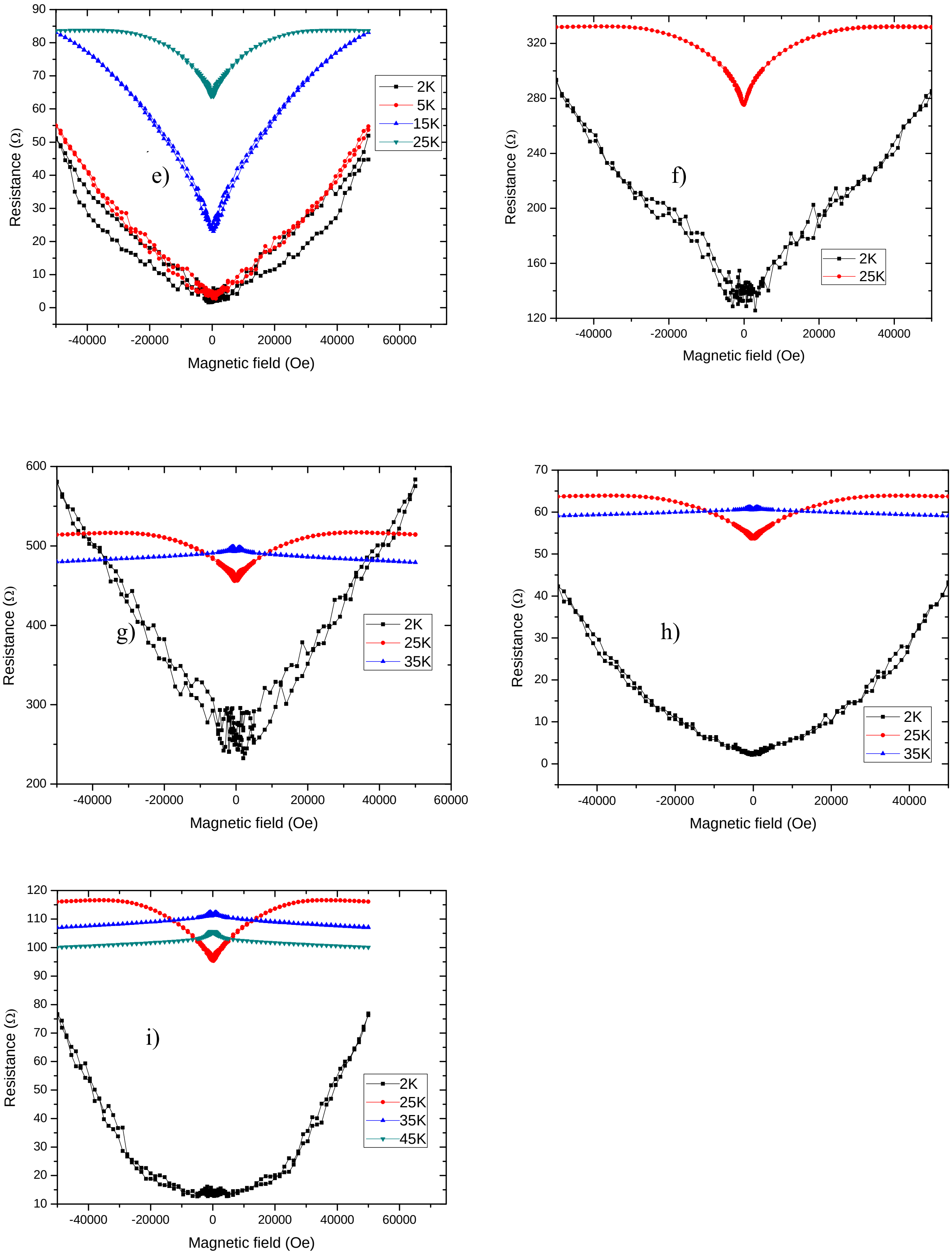

Fig. 11 Magnetic field dependent resistance of $(CrO_2)_x(MgB_2)_{1-x}$ cold-pressed samples for $x$ = 0, 0.1, 0.25, 0.29, 0.32, 0.33, 0.34, 0.35, 0.36 in a) – i) respectively.

We note that for the concentration range very close to the threshold shown in the inset in Fig. 5 hysteretic behavior has been repeatedly observed at the lowest measurement temperature (~2K) (Fig. 11 e) – i)). We speculate that this behavior might be related to the vortex motion in $MgB_2$; more systematic studies are needed to confirm this assumption. Qualitatively similar effects have also been observed in the LSMO/$MgB_2$ composite system. In Fig. 12 the magnetic field dependent resistance curves for $(LSMO)_{0.8}(MgB_2)_{0.2}$ are presented, showing the behavior analogous to that for $CrO_2)_x(MgB_2)_{1-x}$ samples (see Fig. 11).

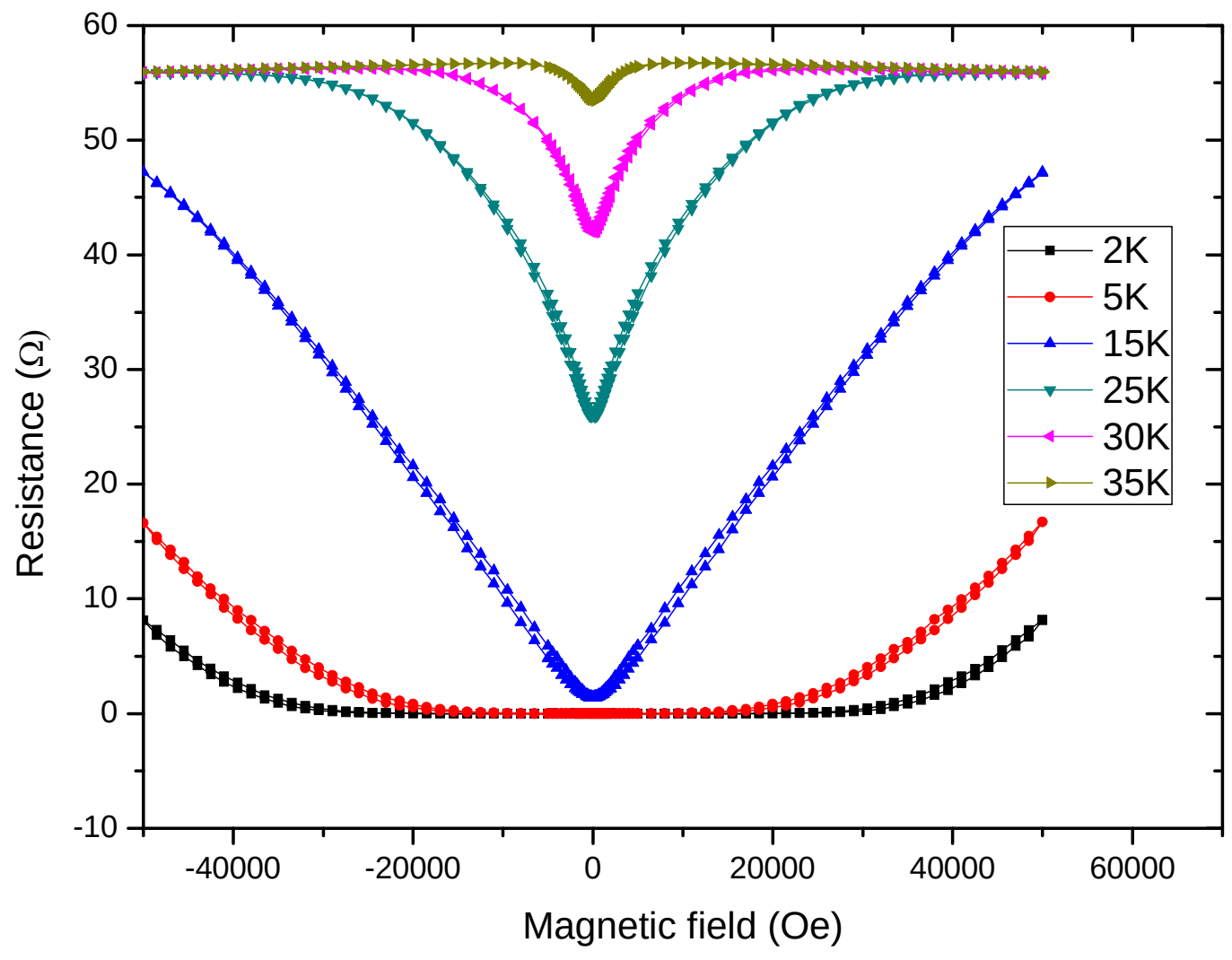

Fig. 12. Resistance vs. magnetic field for $(LSMO)_{0.8}(MgB_2)_{0.2}$ samples at different temperatures.

## IV. CONCLUSIONS

In summary, we have designed a simple two-component percolation system, composed of nanoparticle mixture of a half-metallic ferromagnet ($CrO_2$) and a superconductor ($MgB_2$). The samples with $MgB_2$ concentration below the superconductive percolation threshold show negative magnetoresistance, while positive magnetoresistance is observed above the threshold. The highest magnetoresistance ratio 42% at liquid helium temperatures is obtained for the samples with the largest resistance, at the concentration $x$ = 0.39 close to the percolation threshold.

The behavior of this system is controlled by both its materials properties, such as interface tunneling between the superconducting and ferromagnetic nanoparticles, and by its geometric properties, such as (size/shape) disparity of the constituents. The materials properties of the system are such that direct conductance path between individual particles of different components ($CrO_2$ and $MgB_2$) is effectively blocked, a feature that becomes increasingly prominent at low temperatures. Our study of the transport properties of this system reveals the presence of a double percolative transition, showing a novel phenomenon of conductor-insulator-superconductor crossover effect. The conventional scaling laws of percolation for resistance can be found in both transitions, although with different critical exponents and percolation thresholds. Our simple model can explain the anomalously high threshold observed for the spherical $MgB_2$ particles, which is directly related to the high geometric anisotropy of the constituent particles, resulting in the large exclusion volume of highly anisotropic $CrO_2$ sticks. By replacing $CrO_2$ particles with ferromagnetic and approximately spherical LSMO particles we found that the percolation threshold $p_c$ is much smaller (in the range between 0.32 and 0.39), confirming this assumption.

Although the focus of this work is on the ferromagnet/superconductor nanocomposites, our results can be applied to other multi-component systems showing similar transport properties, as long as the double percolation conditions identified above are satisfied. This can then provide a new avenue for understanding the nontrivial connection between material property/functionality and the structural/geometric behavior of its constituents, which is especially important for multi-component systems. Another implication of this study is the possibility of adjusting/controlling system properties via varying geometric factors (size, shape, anisotropy, etc.) of the constituents, particularly to realize different system modes that are important in material applications (e.g., conducting/insulating/superconducting modes as given in this paper). Specifically, one may be able to adjust one (or both) of the percolation thresholds by varying the size of one of the constituents to bring the non-percolative region in the composition-resistance phase diagram almost to a point, so that the resistance of such a system will become essentially bi-modal, a useful property for various device applications.

## V. ACKNOWLEDGEMENT

We thank Ram Seshadri and Alex Gurevich for discussions and many useful suggestions. The work at Wayne State is supported by the National Science Foundation (CAREER ECS-0239058 and DMR-1006381) (B. N.) and CAREER DMR-0845264 (Z.-F.H.). The UCSB-LANL Institute for Multiscale Materials Studies, the National Science Foundation (DMR 0449354), and the use of MRL Central Facilities, supported by the MRSEC Program of the NSF (DMR05-20415), a member of the NSF-funded Materials Research Facilities Network (www.mrfn.org) are gratefully acknowledged (D.S.).

-------------------------------------

*Present address: Materials Science Division, Argonne National Laboratory, National Laboratory, Argonne*, IL 60439.*